\documentclass[12pt,letterpaper,preprint]{aastex}
\usepackage{graphicx,amssymb,amsmath}
\usepackage{booktabs}
\usepackage{ulem,color}
\usepackage{natbib}
\usepackage{underscore}
\usepackage[flushleft]{threeparttable}

\begin{document}

\author{Julian Krolik$^1$, Tsvi Piran$^2$,  Gilad Svirski$^2$,  and Roseanne M. Cheng$^1$}
\affil{1. Physics and Astronomy Department, Johns Hopkins University, Baltimore, MD 21218, USA
\\
2. Racah Institute of Physics, The Hebrew University of Jerusalem, Jerusalem 91904, Israel}

\title{ASASSN-14li: A Model Tidal Disruption Event}

\begin{abstract}

ASASSN-14li is a recently-discovered tidal disruption event with an exceptionally rich data-set:
spectra and lightcurves in soft X-rays, UV, optical, and radio.    To understand its emission properties in all these
bands, we have extended our model for post-tidal disruption accretion and photon production  to estimate
both soft X-ray radiation produced by the ``prompt" accretion phase and synchrotron emission associated with
the bow shock driven through an external medium by the unbound tidal debris, as well as optical and UV light.
We find that fiducial values of the stellar mass ($1 M_\odot$) and black hole mass ($10^{6.5} M_{\odot}$) yield:
quantitative agreement with the optical/UV luminosity, lightcurve, and color temperature; approximate agreement
with the soft X-ray spectrum and lightcurve; and quantitative agreement with the radio
luminosity, spectrum and lightcurve.  Equipartition analysis of the radio data implies that the radio-emitting
region expands with a constant speed, and its magnitude is comparable to the speed expected for the unbound
stellar ejecta. Both facts provide  strong support to our model.  We find that the disruption event took place
in mid-September 2014.  Two independent parameters, the magnitude and logarithmic radial gradient of the ambient
gas density near the black hole, must be fit to the data to explain the radio emission; their inferred values are
comparable to those found near both Sgr A* and the TDE candidate {\it Swift }~J1644.

\end{abstract}

\section{Introduction}\label{sec:intro}

In recent years, tidal disruptions of stars by supermassive black holes have moved from being merely the object of theoretical speculation to the subject of serious observational investigation \citep[e.g.][]{Komossa+2004,Gezari+2009,Bloom+2011,Gezari.et.al.2012,Chornock+2014,Holoien+2014,vanVelzen2014,Arcavi+2014,Cenko+16}.   However, as perhaps should not be surprising, in doing so they have revealed to us a number of problems in our preconceived notions of how these events behave.

According to the picture of tidal disruptions that has prevailed for the last few decades \citep{Rees.1988,Phinney.1989}, immediately after a tidal disruption the remains of the star are flung out on highly elliptical orbits, with roughly half of the mass actually unbound.   When the bound portion returns to pericenter, it has been thought that relativistic apsidal precession wraps the tidal debris streams so strongly that they shock against each other near the black hole with speeds of order the local orbital velocity.   As a result, they would speedily dissipate a sizable fraction of their orbital energy and acquire nearly-circular orbits, forming a more-or-less conventional accretion disk whose outer edge lies at roughly twice the star's pericenter distance from the black hole.   In such a small accretion disk, the inflow time is rapid, implying that the mass accretion rate onto the black hole---and therefore the bolometric luminosity---should reflect the rate at which mass returns after tracing a single elliptical orbit, a rate expected to decline from its peak $\propto t^{-5/3}$. 

Before evaluating the success of this model, a summary of the various observational predictions based upon it is necessary.  The simplest version, which is the one most often invoked to interpret observations, is that the luminosity in all bands should follow the $t^{-5/3}$ time-dependence of the mass-return.   Others (e.g., \citet{Lodato2011}) apply classical thin accretion disk theory to argue that the bolometric output should follow the mass-return rate closely, but emerge predominantly in the soft X-ray band; the optical light would then be both a very small fraction ($\lesssim 10^{-4}$) of the bolometric luminosity and decrease more slowly with time because, at the outer rim of a disk extending to only twice the stellar pericenter, the temperature is still considerably above the energy of optical photons.  In this version, the optical spectrum would, at early times, be entirely in the Rayleigh-Jeans regime, gradually softening late in the event.  On the other hand, it has also been argued that because the peak mass-return rate can easily be well above Eddington, photon trapping may limit the peak bolometric luminosity to at most a few times the black hole's Eddington luminosity \citep{Loeb.Ulmer.1997,Krolik.Piran.2012}.  This version (as used, for example, in the TDEfit model of \citet{Guillochon.Manukian.Ramirez-Ruiz2014}) would, of course, predict a long period of constant bolometric luminosity and shift its characteristic energy downward toward the EUV; even in the optical band, the luminosity would be fixed until late times.  In a further variation, \citet{Strubbe.Quataert.2009} and \citet{Metzger.Stone.2015} have suggested that the harder photons emitted by the disk might be degraded into the optical band by reprocessing in the unbound debris or a radiation-driven wind; in the model of \citet{Strubbe.Quataert.2009}, the reprocessed fraction could be as much as $\sim 1/3$ for $\sim 10^6 M_\odot$ black holes, but far less for more massive black holes, $\sim 10^{-2}$ for $\sim 10^7 M_\odot$ black holes.   However, in this case the optical luminosity would no longer be $\propto t^{-5/3}$ and its color temperature should increase with time.

Despite the wide range of variations advocated, all of the predictions about the optical light face difficulties when confronted with obsevations.   Observed optical luminosities are typically $\sim 10^{43}$--$10^{44}$~erg/s, $\sim 0.1\times$ the Eddington luminosity for black holes $\sim 10^6 M_{\odot}$, but the observed spectra are frequently fit by black bodies of constant temperature \citep{Gezari.et.al.2012,Chornock+2014,Holoien+2014,vanVelzen2014,Arcavi+2014}, inconsistent with either a disk or an expanding reprocessor origin.  The time-integrated optical light energy is not even as large as the energy that must be dissipated in shocks if the flow is to form an accretion disk on the tidal radius scale.  Moreover, the optical lightcurves actually do sometimes follow roughly the expected $t^{-5/3}$ decline, again contrary to all the elaborations of the classical model except the very simplest and least physical.

Still more discrepancies arise in the details of the spectra.  Although other accreting supermassive black holes (Seyfert galaxies, quasars) uniformly display strong hard X-ray emission in addition to thermal peaks in the UV, this spectral component is quite rare in tidal disruptions: to date, hard X-rays have been seen in only three examples: two objects (Swift J1644+57 and Swift J2058.4+0516) generally thought to be dominated by jets beamed in our line of sight \citep{Burrows11,Bloom+2011} and one which is, in other respects, apparently thermal \citep{Cenko+12}.   In addition to these surprises in the continuum, there are also spectral lines in the optical and ultraviolet whose line widths are $\sim 10^3$--$10^4$~km~s$^{-1}$\citep{vanVelzen+2011,Gezari.et.al.2012,Arcavi+2014,Cenko+16}, indicating an origin $\sim 10^3$--$10^5r_g$ from the black hole, roughly two orders of magnitude farther than the tidal radius if the line widths are interpreted as due to orbital motion.

Detailed numerical hydrodynamics simulations  \citep{Shiokawa+2015} have also cast doubt on the assumptions of the conventional model, demonstrating that the shocks encountered by most of the star's bound mass are too weak to permit the material to join a small-radius accretion disk.    Only if the star's pericenter is exceptionally close to the black hole does the relativistic apsidal precession create stream self-intersections at small enough radius to dissipate a significant fraction of the orbital energy \citep{Dai+15,Sadowski+15}.

Recently, a particularly interesting example, ASASSN-14li\footnote{ASASSN is an acronoym for All-Sky Automated Survey for Supernovae.}, has been found \citep{ATEL14li}.   Discovered in an optical monitoring survey, lightcurves and spectra in the optical and near UV \citep{Holoien+16,Cenko+16}, soft X-rays \citep{Holoien+16,Miller+15,Charisi+16}, and radio \citep{Alexander+16,vanVelzen+2015} have all been obtained.
   
ASASSN-14li's optical lightcurve has been variously described as exponential with a decay time of $\simeq 60$d \citep{Holoien+16} or $\propto (t-t_*)^{-5/3}$ with $t_* \simeq 35$d before the 22~November~2014 discovery \citep{Miller+15}.   Spectra indicate it maintained a nearly-constant color temperature $\simeq 3.5 \times 10^4$~K over a period of several months \citep{Holoien+16,Cenko+16}.   The maximum observed optical luminosity was $\simeq 2.5 \times 10^{43}$~erg~s$^{-1}$ \citep{Holoien+16}; if its spectrum is truly a blackbody with the measured color temperature, the peak total optical/UV luminosity was $\simeq 6 \times 10^{43}$~erg~s$^{-1}$ \citep{Holoien+16}.  

The first X-ray observations were less than a week after discovery.  The flux reached a peak 20--30~d after discovery \citep{Miller+15,Charisi+16}, at which its inferred bolometric luminosity was $\simeq  3 \times 10^{44}$~erg~s$^{-1}$; at later times, its flux declined approximately linearly at first, but with the slope later becoming shallower.  However, the $1\sigma$ uncertainty in individual data points is $\simeq 1/4$ of the entire drop in flux, so little can be said about finer details of the lightcurve.
Like the optical light, the color temperature remained roughly constant over time at a temperature 6--$7 \times 10^5$~K \citep{Miller+15,vanVelzen+2015}.  Thus, the soft X-rays are the single largest contributor to the total luminosity, and, at least for the first half-year or so after discovery, during which most of the event's energy was radiated, the soft X-ray lightcurve was not a power-law at all, much less $\propto t^{-5/3}$.

Numerous line features have also been seen in its spectrum, in both the optical/UV and the soft X-ray bands.   In the optical and UV, there are emission lines with widths $\sim 10^3$--$10^4$~km~s$^{-1}$ and absorption lines of considerably smaller width, generally a few hundred km/s \citep{Holoien+16,Miller+15,Cenko+16}.   The absorption lines in both the UV and X-rays are also blue-shifted by a few hundred km/s {while the emission lines appear to be approximately centered on the host galaxy redshift} \citep{Miller+15,Cenko+16}.

Two different groups monitored the flare's radio flux.   Both find that the high frequency ($> 10$~GHz) flux declines steadily from 1 month after discovery to 10 months later, and both also agree that the total radio spectrum from $\simeq 1$~GHz to $\simeq 20$~GHz was initially fairly flat, but gradually becomes steeper.   Nonetheless, their interpretations are quite different: \citet{Alexander+16} attribute much of the low-frequency flux to a pre-existing time-steady and optically-thin synchrotron source, whereas \citet{vanVelzen+2015} argue that the pre-existing source is suppressed and then replaced by a jet associated with the TDE.   
In the \citet{Alexander+16} picture, the optical/UV light of the disruption flare drives a large opening-angle outflow at 12,000~km~s$^{-1}$ beginning $\simeq 90$~d before optical discovery.   The contrasting model of \citet{vanVelzen+2015} relies on the observations of \citet{Falcke+2000} to argue that the pre-existing radio source was very compact, so that it could be entirely suppressed by the TDE.   On this basis, they argue that the flare radio emission is optically thin, and therefore so far from the disrupting black hole that it must be due to a relativistic jet launched as a result of the tidal disruption.

With such a complete dataset, as well as such contradictory interpretations, ASASSN-14li invites further analysis.    In the remainder of this paper we show how essentially all its remarkable properties are described quite well by a new way of looking at TDEs, partly described in the work of \citet{Shiokawa+2015}, \citet{Piran+2015}, and \citet{Svirski+15}, and further developed here. A remarkable feature of this new approach is that it does not involve any ad hoc components (such as jets or outflows driven by super-Eddington radiation forces) invoked to explain a particular spectral component; it is based solely on the hydrodynamics of the tidally disrupted stellar matter.

\section{Theoretical Predictions}\label{sec:theory}

\subsection{Overview}

At the beginning of a tidal disruption event, a star of mass $M_*$ and radius $R_*$ follows an essentially parabolic orbit with a pericenter $R_p$ from the black hole.   If $R_p \leq R_T \sim R_* (M_{\rm BH}/M_*)^{1/3}$, once the star passes within $R_T$, the further trajectories of its material are roughly described by Keplerian orbits in the black hole potential.   These orbits are characterized by the specific angular momentum of the star and the specific energy of individual fluid elements within the star when it is near the tidal radius: matter on the near side has specific energy ${\cal E} \simeq -GM_{\rm BH}R_*/R_T^2$, matter at the center has ${\cal E} \simeq 0$, matter on the far side has ${\cal E} \simeq +GM_{\rm BH}R_*/R_T^2$ \citep{Stone.Sari.Loeb2013}.    This range of energies implies that roughly half the mass is unbound, while the bound half traverses orbits with semi-major axes ranging from $a_{\rm min} \sim R_T(M_{\rm BH}/M_*)^{1/3}$ to infinite.    The orbital period for an orbit with semi-major axis $a_{\rm min}$ determines a characteristic evolutionary timescale for the system $t_0$.    These orbits are all highly eccentric, with eccentricity $e \simeq 1 - 2 (M_*/M_{\rm BH})^{1/3}$; if the tidal debris is ever to form an approximately circular accretion disk at $\sim R_T$ it must lose a large amount of energy, of order $G M_* M_{BH} /R_T$, which is not that much smaller than typical accretion energy release because  $R_T$ is only tens of $r_g$ (see eqn.~\ref{eq:RT}).

\cite{Shiokawa+2015} presented a detailed numerical simulation of TDE hydrodynamics that followed the debris until nearly all of the bound mass had traveled out to apocenter and returned to the vicinity of the black hole.   When the first streams of formerly stellar mass return to the black hole, their convergence toward the orbital midplane creates a shock near the pericenter.   Predicted by early analytic efforts \citep{Evans.Kochanek.1989,Kochanek1994}, this shock is often called the ``nozzle" shock.    In this shock, the streams converge at almost glancing incidence, making the shock speed a small fraction of the orbital speed.   The energy dissipated in it is therefore only a very small fraction of the amount necessary to ``circularize" the streams' orbits.    Having passed through the nozzle shock, the streams head outward; near their apocenter they intersect with newly-arriving stellar matter, debris with somewhat larger orbital semi-major axis and therefore orbital period.    A pair of shocks is created at that intersection, one in the newly-arriving matter, and one in the matter that has already gone around the black hole at least once.    Meanwhile, a pile-up of material at the nozzle shock lengthens its front in the radial direction, stretching it both inward and outward, but also decreasing its Mach number.    As a result, subsequent streams encountering this shock suffer greater deflection, but dissipate less energy.    Beginning at a time $\simeq (2$--$3)t_0$, a fraction of the matter encountering the nozzle shock is pushed sharply inward.    This matter has such small specific angular momentum that it can accrete onto the black hole in a time short compared to $t_0$, maintaining an accretion rate that is roughly constant at a rate $\lesssim 0.1\times$ the maximum mass return rate until $\simeq 8$--$10t_0$.     About $ 25$--$30\%$ of the bound mass is accreted in this fashion.

Almost simultaneously with the beginning of matter flow to small radii, the total shock heating rate rises.    Initially the greatest heating takes place at the inner nozzle shock, where the heating rate peaks at $\simeq 3t_0$ and then falls to almost nothing by $\simeq 6t_0$.   For times later than $\simeq 5t_0$,  the outer shocks dominate, with a heating rate comparable to that of the nozzle shock at its peak.     The heating rate in the outer shocks peaks at $\simeq 7t_0$, and thereafter drops $\propto t^{-5/3}$, as it is largely dependent upon the arrival of new material with still greater orbital periods.   As already noted in \citet{Piran+2015}, the outer shock energy dissipation rate agrees quite well with the observed luminosity, time-dependence, and effective temperature of the optical/UV radiation seen in TDEs.

As a result of these complicated hydrodynamical interactions, the majority of the star's bound mass is spread over a range of radii comparable to the semi-major axis of the most tightly-bound material.    Even after more than $10t_0$, the surface density of this matter remains highly asymmetric and irregular.   The shocks heat the gas sufficiently to make the flow geometrically thick; the ratio of its density scale-height to radius $H/R \simeq 0.4$, almost independent of radius.  Scaling the simulation data to the parameter values expected in typical events (main sequence stars of mass $M_* \sim 1 M_{\odot}$ and black holes of mass $M_{\rm BH} \sim 10^{6.5} M_{\odot}$) suggests that the local cooling time is larger than the local orbital time out to radii $\sim a_{\rm min}$ \citep{Piran+2015}.\footnote{We deem this black hole mass ``typical" on the grounds that the black hole in our own galaxy has a mass $\simeq 4 \times 10^6 M_{\odot}$, while \citet{Miller+15} estimate that the black hole mass in ASASSN-14li is $\simeq 2.5 \times 10^6 M_{\odot}$ if the peak X-ray luminosity was exactly Eddington, and \citet{vanVelzen+2015} estimate $\simeq 6 \times 10^6 M_{\odot}$ from the host galaxy's bulge/total luminosity ratio.}    Because the mean specific angular momentum of the debris is not that much larger than the critical value at which matter can pass through the ISCO, it is possible that internal MHD stresses within the flow may lead to accretion of most of the matter {\it without} dissipating a sizable part of its orbital energy \citep{Svirski+15}; the shortest plausible time required to accrete most of the mass by this mechanism is $\sim 10t_0$.

Meanwhile, the unbound material never returns to the black hole, coasting outward at a speed $\sim [GM_{\rm BH}/a_{\rm min}]^{1/2}$.  To zeroth order, the unbound material is confined to a thin wedge spanning the stellar orbital plane whose opening angle is $\sim R_*/R_T \sim (M_*/M_{\rm BH})^{1/3}$.   
However, once it reaches radii $\gtrsim a_{\rm min}$, which it does $\simeq t_0/2$ after the disruption, the bow shock it drives in the ambient gas raises the shocked material to temperatures above the local virial temperature.   As a result, this gas rapidly expands away from the ejecta orbital plane; as it does,  the bow shock stretches farther behind the leading edge of the ejecta as well as farther away from the orbital plane, reaching an opening angle  $\sim (M_*/M_{\rm BH})^{1/9}$; as we explain in more detail in Sec.~\ref{sec:ejecta}, the expanding-wedge geometry of the ejecta makes the bow shock opening angle scale with the 1/3 power of the wedge opening angle. In effect, this sideways expansion of the shocked ambient gas creates a system resembling a supernova remnant \citep{Guillochon+15}.    The bow shock surrounding the unbound ejecta can both accelerate electrons to relativistic velocities and amplify the ambient magnetic field, producing a synchrotron-radiating region.

\begin{figure*}[!ht]
\centering
\includegraphics[width=0.75\textwidth]{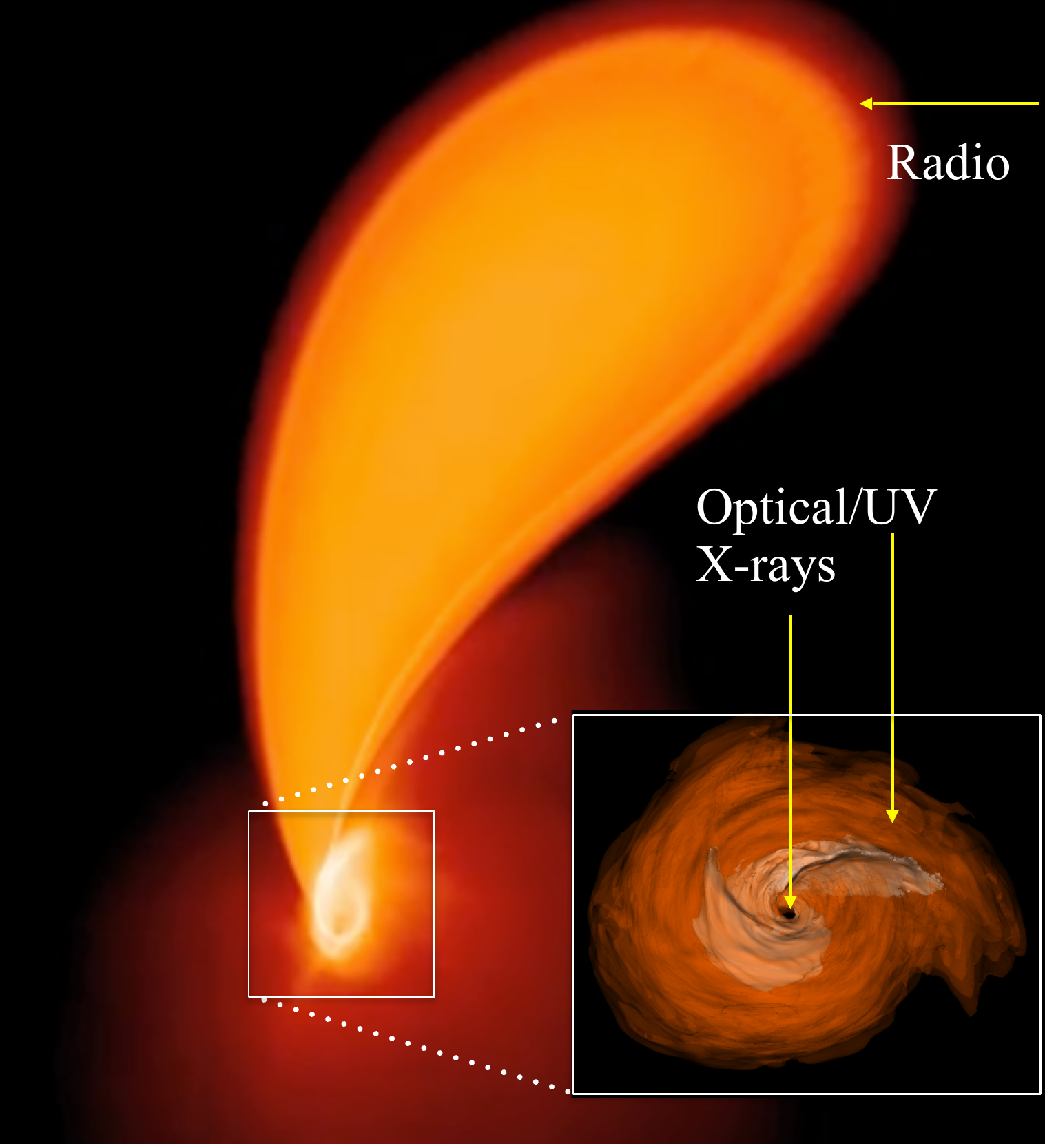}
\caption{
The basic components of our model.   The late time density profile of the stellar debris from a simulation by \citet{Rosswog.et.al.2009} is shown in white/gold/red color in the large panel (see \texttt{http://compact-merger.astro.su.se/Movies/IMBH1000_WD02_4e6parts_P12_N.mov}).  The late time internal energy profile of the debris is shown in the inset (data from \citet{Shiokawa+2015}); this quantity highlights shock locations.   In our model the observed optical/UV radiation arises from the interaction of freshly infalling matter with a cloud of matter formed around the SMBH at $R \sim a_{\rm min}$, the scale of the inset.  Radio emission arises much farther out from interaction of the unbound ejecta with the surrounding matter at $R = v_o t$.   X-rays are radiated much closer to the black hole, from the nozzle shock at $R \sim R_p$ inward toward the ISCO and event horion.}
\label{fig:ba}
\end{figure*}

We propose that the salient properties of this picture: the shocks in the apocenter region, the outflow of unbound material, and the matter directed promptly to the black hole by the nozzle shock very naturally explain, respectively, the optical/UV emission, radio emission, and soft X-rays observed in the case of ASASSN-14li.   These  basic components of the model are depicted in Fig.~\ref{fig:ba}. Note that because of optical depth effects particularly important to optical/UV and X-ray radiation, photons emerge from the same general region where their energy was generated, but a fully-resolved image of the system would not correspond in detail with a map of dissipation.

\subsection{Specific predictions}

\subsubsection{Basic parameters}

These dynamical results lead to a number of observational predictions of direct relevance to ASASSN-14li.
To make these connections, we need to attach physical scales to basic parameters, beginning with the tidal radius.
Using the same fiducial parameters chosen in \cite{Piran+2015}, the tidal radius in gravitational units ($r_g = GM/c^2$) is
\begin{equation}\label{eq:RT}
R_T \simeq 15 [(k/f)/0.08]^{1/6} (M_*/M_{\odot})^{2/3-\xi} M_{BH,6.5}^{-2/3} r_g.
\end{equation}
Here $k/f$ parameterizes structural properties of the star so that the tidal radius $R_T = (k/f)^{1/6} R_* (M_{\rm BH}/M_*)^{1/3}$ \citep{Phinney.1989}; it ranges from 0.02 for fully-radiative stars to 0.3 for fully-convective stars, so we have scaled to the geometric mean of these two extremes.   We describe the main sequence mass-radius relation by $R_* = R_{\odot} M_*^{1-\xi}$, where $\xi \simeq 0.2$ for $0.1 < M_* < 1$, but rises to $\simeq 0.4$ for $1 < M_* < 10$ \citep{Kippenhahn1994}.

The characteristic timescale corresponding to the orbital period of the most bound matter is
\begin{equation}
t_0 \simeq 20 [(k/f)/0.08]^{1/2} M_{\rm BH,6.5}^{1/2} (M_*/M_{\odot})^{(1-3\xi)/2}\hbox{~d},
\end{equation}
while the peak accretion rate onto the black hole, which the simulation of \citet{Shiokawa+2015} showed to be $\simeq 0.1\times$ the peak mass-return rate, is
\begin{equation}
\dot m_{\rm early} \simeq 6 (\eta/0.1) \left({k/f \over 0.08}\right)^{-1/2} M_*^{(1+3\xi)/2} M_{BH,6.5}^{-3/2}
\end{equation}
when measured in Eddington units assuming a nominal radiative efficiency $\eta$.  Provided the black hole mass is not too large nor the stellar mass too small, $\dot m_{\rm early}$ can be significantly super-Eddington.

\subsubsection{Optical/UV emission from the outer shocks}

As discussed in \citet{Piran+2015}, the shocks near the streams' apocenters produce radiation primarily in the optical/UV band.   Beginning at $\simeq 3$--$4t_0$, this outer shock heating reaches a peak at $t \simeq 7t_0$.  \citet{Piran+2015} estimate a photon diffusion time in this region $\simeq 3t_0 M_{\rm BH,6.5}^{-7/6}$; if the heat is all radiated promptly, at least relative to the $t_0$ timescale, the maximum optical/UV luminosity is
\begin{equation}
L_{\rm opt} \simeq 8 \times 10^{43} [(k/f)/0.08]^{-5/6} (M_*/M_{\odot})^{1/6 + 5\xi/2} M_{BH,6.5}^{-1/6}\hbox{~erg~s$^{-1}$}.
\end{equation}
The fact that the photon diffusion time is comparable to or somewhat greater than the orbital period has the consequence that the flow stays warm even when matter has traveled to the opposite side of the flow from the place where it was shocked.    For this reason, we expect the flux from the flow surface to be spread more widely than the shocks responsible for its energy.

At later times, the optical/UV luminosity should decline $\propto t^{-5/3}$ because the heating supporting it is derived directly from the shocks created when newly-returning matter encounters the outer edge of the accretion flow.    Because the geometrical thickness of the gas streams is small at radii outside the shocks, the optical/UV light should be seen readily from almost all directions.

The peak effective temperature at the flow's photosphere near the outer shocks may be as high as $\simeq 5 \times 10^4[(k/f)/0.08]^{-3/8} (M_*/M_\odot)^{-1/8 + 9\xi/8} M_{\rm BH,6.5}^{-3/8}$~K \citep{Piran+2015}.   One would therefore expect to see blue colors in the optical/UV region, but with spectral slopes somewhat shallower than the Rayleigh-Jeans limit of $F_\nu \propto \nu^2$.    There might also be a small amount of spectral curvature.

Although the density and optical depth
($n \sim 6 \times 10^{12} [(k/f)/0.08]^{-1}(M_*/M_{\odot})^{3\xi} M_{\rm BH,6.5}^{-2}$~cm$^{-3}$, $\tau \sim 500 [(k/f)/0.08]^{-2/3} (M_*/M_{\odot})^{1/3+2\xi} M_{\rm BH,6.5}^{-4/3}$) of the material near the outer shocks are large enough to produce a thermalized spectrum, line features would not be surprising.    In fact, these conditions are quite similar to those in the reprocessing atmosphere calculation of \citet{Roth15}, which predicts a number of emission lines.   The principal contrast is that the continuum flux at the bottom of the atmosphere is somewhat smaller in our situation, and it is possible that there may be some external illumination by X-rays from the central part of the accretion flow (as discussed in Sec.~\ref{sec:X-rays}).  If any emission lines do form, their full  width (FW0I) would be comparable to the spread in fluid velocities.   For matter at a distance $a_{\rm min}$ from the black hole, this is a line-of-sight projection factor times $\sim 22,000[(k/f)/0.08]^{-1/6} M_{\rm BH,6.5}^{1/6} (M_*/M_{\odot})^{-1/6+\xi/2}$~km~s$^{-1}$ (twice the circular orbit speed at that radius).   The line profile should be reasonably well centered on the galaxy rest-frame because of the rough azimuthal symmetry in the flow's surface brightness.

\subsubsection{Radio emission from shocks driven by unbound tidal debris}\label{sec:ejecta}

At radii beyond $a_{\rm min}$, the unbound matter moves outward at a characteristic speed that is similar to the characteristic orbital speed at $a_{\rm min}$ because the energy distribution of the tidal debris is symmetric about zero.
The speed $v_{\rm out}$ is the speed the outflow reaches ``at infinity", i.e., $R \gg a_{\rm min}$.  For fixed total energy, the unbound debris actually travels faster at smaller radii.  However, because we are mostly interested in times many $t_0$ after disruption, the higher initial speed makes little difference to observational predictions.  This outward speed at $R \gg a_{\rm min}$ is
\begin{equation}
v_{\rm out} \simeq \left( \frac{2GM_{\rm BH}^{1/3} M_*^{2/3}}{(k/f)^{1/3}R_*}\right)^{1/2} \simeq 11,000 [(k/f)/0.08]^{-1/6} M_{\rm BH,6.5}^{1/6} (M_*/M_{\odot})^{-1/6+\xi/2}\hbox{~km~s$^{-1}$},
\end{equation}
so that the flow carries $\simeq 6 \times 10^{50}[(k/f)/0.08]^{-1/3} M_{\rm BH,6.5}^{1/3} (M_*/M_{\odot})^{-1/3+\xi}$~erg in kinetic energy.
Detailed calculations of the energy distribution function of the tidal debris \citep{Cheng.Bogdanovic.2014,Shiokawa+2015} generally show a sharp edge at energies immediately above that corresponding to $v_{\rm out}$.  However, these simulations are not well-suited to determining the extent of any high-energy tail to this distribution.  In part this is because they lack the necessary resolution, and in part because such a tail depends on the star's internal structure, and most simulations to date adopt the polytropic approximation for their initial condition.  Any matter in such a tail will travel faster than $v_{\rm out}$ and form the leading edge of the unbound material. The ejecta continue to move with constant velocity until they sweep up a mass comparable to their own,  which requires traveling very far from the black hole, so far that deceleration begins too long after the TDE to be relevant.   Thus, the radius of the shock at a time $t$ since the stellar disruption is $r = v_{\rm o} t$, where $v_{\rm o}$, the speed of the fastest debris with enough mass to generate the observed radio emission, is $\geq v_{\rm out}$.

The ejecta drive a {(forward)} bow  shock that propagates into the surrounding matter and a reverse shock that propagates much more slowly upstream into the ejecta.   The energy dissipation per nucleon in the forward shock  is  $\sim  m_p v_{\rm o}^2$. A fraction $\epsilon_e$ of this energy goes to accelerate electrons, while  a fraction $\epsilon_B$  is used to generate a magnetic field, so that  the field is
$B = [16 \pi \epsilon_B n(r) m_p v_{\rm o}^2 /2]^{1/2} $.  Here $n(r)$ is the external density, and we assume that the surrounding matter has a density profile $n(r) = n_0 (r/r_0)^{-\hat k} $. Using the measured Galactic Center gas density distribution at slightly larger radius, $n \propto r^{-1}$, with $n=130$~cm$^{-3}$ at 0.04~pc \citep{Baganoff.et.al.2003}, we set our fiducial density to  $n_0\approx 1500$~cm$^{-3}$ at $r_0=10^{16}$~cm.
A comparable density has been inferred in gas near the TDE candidate Swift~J1644 \citep{BarniolPiran2013}.

Under these conditions, the hot shocked electrons produce synchrotron emission.   The region is optically thick to self-absorption \citep[see e.g.][]{Pacholczyk70,Chevalier98} at frequencies below the self absorption frequency: 
\begin{eqnarray}
\nu_a = 0.4 f_A^{-2/7}  f_V ^{2/7}& (\epsilon_e/0.1)^{2/7}  (\epsilon_B/0.1)^{5/14} (\gamma_m/2)^{2/7} (n_0/1500\hbox{~cm$^{-3}$})^{9/14} (r_0 /10^{16}\hbox{~cm})^{9 \hat k/14} \\ \nonumber
        &(t /100\hbox{~d})^{(4 - 9 \hat k)/14} (v_{\rm o}/ 11,000 {\rm km/s})^{(14 - 9 \hat k)/14}\hbox{~GHz} \ .
\label{nua}
\end{eqnarray}
 The factors  $f_A$  and $f_V $ are defined so that the emitting region has an area $ f_A \pi R^2$ and a volume $f_V  \pi R^3 $.\footnote{These definitions correspond to those of  \cite{Barniol+13} who considered relativistic shocks with a Lorentz factor $\Gamma$, for which the emitting area is  $ f_A \pi R^2/\Gamma^2 $ and the emitting volume is $f_V  R^3/\Gamma^4$ . In the fully isotropic Newtonian case, $f_A=4$ and $f_V = 4 \pi/3$ . }
The corresponding flux at a distance $d$ is
\begin{eqnarray}
F_\nu(\nu_a)= 3.4 ~f_A^{2/7} f_V^{5/7} 
(\epsilon_e /0.1)^{5/7}  (\epsilon_B/0.1)^{9/14} (\gamma_m/2)^{5/7} (n_0/1500\hbox{~cm$^{-3}$})^{19/14} (r_0 /10^{16}\hbox{~cm})^{19 \hat k/14} \\ \nonumber
\times (t /100\hbox{~d})^{19(2- \hat k)/14} (v_{\rm o}/ 11,000\hbox{~km/s})^{(56- 19\hat k)/14} (d / 2.7 \times 10^{26}\hbox{~cm})^2\hbox{~$\mu$Jy} \ .
\label{Fnua}
\end{eqnarray}
The optically thin spectrum above $\nu_a$ is  a power-law with a slope $-(p-1)/2= -1$. The optically thick flux below $\nu_a$  has the common optically thick synchrotron slope of $+5/2$.  These fiducial values give a marginally detectable signal. However, a higher density or a faster outflow (it is only necessary for a small fraction of the mass to escape at higher speed) would result in a much stronger signal.

These estimates are all posed in terms of a quasispherical expansion, but the unbound debris rush out from the star within a wide, but thin wedge.    Although  the spread in azimuthal angle $\Delta\phi \sim 1$, one might roughly estimate the one-sided spread in polar angle $\Delta\theta$ to be much smaller, only $\sim R_*/R_T \sim (M_*/M_{\rm BH})^{1/3} \sim 10^{-2}$.   However, the radio-emitting electrons are predominantly found in the shocked ambient gas, and its geometry is quite different from the ejecta wedge.    Dimensional analysis in the spirit of the Sedov-Taylor solution suggests the shape of the bow shock surrounding the ejecta.    When the ejecta have reached radii well past $a_{\rm min}$, so that the post-shock temperature exceeds the virial temperature, there are only three dimensional quantities relevant to the vertical extent of the expanding shocked gas: $dE/dS$, the energy injected by shock dissipation per unit area { in the ejecta orbital plane}, the external mass density $m_p n$, and the time $t^\prime$ since the gas was shocked.    There is only one way to combine these quantities to form a distance:
\begin{equation}
z \sim \left(dE/dS\right)^{1/3} \left(m_p n\right)^{-1/3} {t^\prime}^{2/3}.
\end{equation}
If the shocked material has an adiabatic index of 5/3, $dE/dS \simeq (9/16) m_p n v_{\rm o}^2 R \Delta\theta $.  Combining this with the kinematic relation $t^\prime = (R_s - R)/v_{\rm o}$, where
$R_s$ is the current position of the shock and $R$ is the radius at which the gas that has spread to $z$ was shocked, we find that
\begin{equation}
z \sim  \left(\Delta \theta \right)^{1/3} \left( R/R_s \right)^{1/3} \left(1- R/R_s \right)^{2/3} R_s \ .
\end{equation}
Note that because of the planar geometry of the ejecta, this bow shock is wider than the parabolic one obtained for a round obstacle \citep{AlmogReem15}. The half-opening angle is:
\begin{equation}
\Delta\theta_S \sim \left(\Delta\theta \right)^{1/3} { (R_s/R)^{2/3}} \left(1-R/R_s\right)^{1/3}
                \sim { 0.2} {(R_s/R)^{2/3}}\left(1-R/R_s\right)^{1/3} (M_*/M_{\odot})^{1/9} M_{\rm BH,6.5}^{-1/9}.
\end{equation}

Because we estimate the flux at $\nu_a$, where the synchrotron emission is marginally optically thick, the most important aspect of the region's geometry is the area it occupies in the observer's sky plane. If the angle between the line of sight to the observer and the ejecta's orbital plane (which might be different from the stellar orbital plane if the black hole has significant spin) is $\psi$, we can approximate the area covering factor by
\begin{equation}
f_A \simeq \left(2 \sin\psi \, \Delta \theta_S + \cos\psi \right) \Delta\phi/(2\pi).
\end{equation} 
If the ejecta plane is close to the sky plane, we see a wedge whose angular width is $\sim \Delta \phi \sim 1 $; if the ejecta plane is nearly perpendicular to the sky plane, we see a wedge with width $\sim 2\Delta \theta_S$.     The typical scale expected for $f_A \sim 0.2$.

Note that this estimate is not significantly influenced by self-gravity of the unbound debris.    Although self-gravity may initially restrict vertical expansion \citep{Kochanek1994,Guillochon.Manukian.Ramirez-Ruiz2014}, it ceases to be significant at radii too small to affect stream geometry in the radio-emitting region. This is especially clear for the very small fraction of the matter that, as we will show in \S~\ref{sec:radiodata}, is responsible for the observed radio emission in ASASSN-14li. The ejecta driving the shock are, of course, the fastest-moving portion.  By definition, this gas has the most positive net energy, so it came from the outermost layers of the star on the far side of the black hole.  These layers are relatively cool while still in the star, so even a small amount of adiabatic expansion causes their temperature to fall to the level at which hydrogen recombines.   The injection of entropy associated with recombination puts this matter on an adiabat with enough heat content to make self-gravity no longer important \citep{Kochanek1994}.  Similarly, early clumping \citep{Coughlin+16} may influence the small-radius behavior of the outflow, but will not affect the bow shock that arises from the interaction of the unbound ejecta with the surrounding matter far from the black hole.

\subsubsection{Soft X-ray emission from the inner flow}\label{sec:X-rays}

Theoretical predictions for X-ray emission are much less certain than for light in other bands because the only hydrodynamical calculations of how tidal debris travels from the pericenter region to the black hole treat the rather special case of a pericenter much smaller than $R_T$ \citep{Haas+2012,Sadowski+15}.   For this reason, the predictions of this section are somewhat more tentative
 than those of the previous two sections.

Although the nozzle shock is too weak to ``circularize" the majority of the star's bound mass \citep{Guillochon.Manukian.Ramirez-Ruiz2014}, the heating associated with it can be significant on the scale of the observed radiation.    Scaled from the data of \citet{Shiokawa+2015}, its peak dissipation rate is $\simeq 5 \times 10^{43} [(k/f)/0.08]^{-5/6} (M_*/M_\odot)^{1/6+5\xi/2} M_{\rm BH,6.5}^{-1/6}$~erg~s$^{-1}$.    This rate of nozzle shock heating lasts from $\simeq 3t_0$ to $\simeq 6t_0$.  About a factor of 6 below the peak soft X-ray luminosity from ASASSN-14li, this heating rate can be securely taken as a lower bound on the total dissipation rate in the portion of the flow within $\sim R_T$.

Deflection at the nozzle shock also transfers angular momentum from some parcels of gas to others; those losing angular momentum move inward radially.    Travel from the nozzle shock to the black hole can happen quickly when measured in units of $t_0$.    If the fluid's specific angular momentum is large enough to form a conventional circular accretion disk, the magneto-rotational instability will drive MHD turbulence to a saturated state in $\sim 10$ local orbital periods, a time that is only $\sim 10 (M_*/M_{\rm BH})^{1/2}t_0$, i.e., $\sim 0.01t_0$ for our fiducial parameters.   Once that is accomplished, the inflow time should be only a few times longer.    Alternatively, if the specific angular momentum is too little to support circular orbits at radii $\sim R_p$, accretion onto the black hole may be even faster, $\sim 10$ local orbits, because the MHD stresses need only to remove enough angular momentum to permit streams to plunge directly across the ISCO \citep{Svirski+15}.

In either case, the flow should be quite hot, with temperature high enough to create a geometrically thick configuration.   In the former case, that of a conventional accretion disk with nearly-circular orbits, the dissipation rate per unit mass should fall in the usual range, implying a total heating rate $\simeq 3 \times 10^{45} [(k/f)/0.08]^{-1/2} (M_*/M_\odot)^{(1+3\xi)/2} M_{\rm BH,6.5}^{-1/2}$~erg~s$^{-1}$ when the accretion rate is in its plateau at $\dot m_{\rm early}$.
 This heating rate corresponds to what is expected in mildly super-Eddington accretion; the flow must then be geometrically thick.  The largest possible soft X-ray luminosity is this heating rate, but, as we are about to argue, the actual luminosity is likely to be rather lower.

 In the latter case, a highly eccentric disk, gas thermodynamics works rather differently than in a conventional disk.   In a conventional disk, the gas's temperature is governed by the balance between local dissipation and local cooling because the inflow time is usually longer than the cooling time.  By contrast, in an eccentric disk, the time for the gas to move radially is the orbital period, which is always shorter than the cooling time.  Consequently, adiabatic processes are relatively much more important to eccentric disks.  In radiation-dominated conditions, the disk temperature $\propto \rho^{1/3}$ when the density changes adiabatically; in an homologous flow, $\rho \propto R^{-3}$; it then follows that $T \propto R^{-1}$, just like the depth of the gravitational potential.   If the temperature at $R_p$ is high enough to make the flow geometrically thick there (i.e, the ratio of radiation pressure to gas density is close to the square of the circular-orbit speed), it stays thick all the way to the ISCO.  As the material follows its orbit inward, gravity does work on the gas, and orbital energy is converted to thermal energy; as the material returns to apocenter, pressure forces return thermal energy to orbital energy.   Thus, in a geometrically-thick eccentric disk, the central temperature reaches levels comparable to those seen in a conventional disk, but achieves such a temperature by different means.    Any dissipation associated with the MHD turbulence only raises the eccentric disk temperature.  Conversely, in a conventional disk, to the degree that slow heat transport prevents full local radiation of local dissipation, the gas follows a higher entropy adiabat as it moves inward and also experiences an adiabatic temperature rise.

The luminosity emerging from such disks is { made still more} uncertain because the photon diffusion time is, { in fact, quite long even if the flow's orbits are circular}.   Scaling from the simulation data of \citet{Shiokawa+2015} at $t \simeq 3t_0$, the local photon diffusion time in the vicinity of the nozzle shock is $t_{\rm diff} \simeq 5 \times 10^6 (M_*/M_\odot)^{1/3+\xi} M_{BH,6.5}^{-1/3}$~s, or $\simeq 3 (M_*/M_\odot)^{-1/6+5\xi/2} M_{\rm BH,6.5}^{-5/6} t_0$.    Although this is similar in absolute terms to the diffusion timescale at $R \sim a_{\rm min}$, it is $\simeq 800 (M_*/M_\odot)^{-2/3+5\xi/2} M_{\rm BH,6.5}^{-1/3}$ orbital periods at $R_p$.  Closer to the black hole, the surface density tends to be a few times smaller, while the scale height $H$ diminishes proportional to the radius.   Thus, { at $R < R_p$}, the diffusion time could be an order of magnitude shorter, but this would remain a similar number of local orbits.  

In the instance of a conventional circular-orbit disk with $H/R \sim 1/2$, the inflow time $t_{\rm inflow}$ is at least several dozen orbits, but that is still considerably shorter than the photon diffusion time.   The luminosity attributable to photon diffusion would then be a fraction $t_{\rm inflow}/t_{\rm diff}$ of the circular-disk heating rate.    More quantitatively, at $R_p$ this timescale ratio is $\simeq 0.05$ for circular accretion with stress-to-pressure ratio $\sim 0.1$ and $H/R \simeq 0.5$. With the assumption that $t_{\rm inflow}/t_{\rm diff}$ is a slow function of radius inside $R_p$, the soft X-ray luminosity escaping by photon diffusion from a circular-orbit accretion flow is
\begin{equation}\label{eq:Lx}
L_X \simeq 1.5 \times 10^{44} [(k/f)/0.08]^{-1/2} (M_*/M_\odot)^{7/6-\xi} M_{\rm BH,6.5}^{-1/6}\hbox{~erg~s$^{-1}$}.
\end{equation}
 There are, however, two reasons why $L_X$ might be larger than this estimate: adiabatic compression will raise the temperature at small radius above the level due to dissipation alone; and magnetically-buoyant bubbles may help carry heat to the surface \citep{Blaes2011,Jiang2014}.

 In the case of an eccentric disk, the effective inflow time is the time for the orbital angular momentum to diminish to the point that the matter can plunge directly through the ISCO.  For an eccentric disk formed in a TDE, this timescale is generically only $\sim 10$~orbits because the angular momentum of the material starts out not much greater than the critical value \citep{Svirski+15}.   For this reason, $L_X$ from an eccentric inner flow could be smaller by a factor of several than for a conventional disk, again with some uncertainty due to magnetic buoyancy heat transport.
Given these uncertainties that might change the estimate of eqn.~\ref{eq:Lx} either up or down by factors of several, in the following we will use the estimate of eqn.~\ref{eq:Lx} as our standard prediction, but one should bear in mind it depends on some uncertain physics.

The spectrum of the emergent radiation can be estimated from the luminosity and the size of the region from which it is radiated.   If the surface temperature inside $R_p$ is uniform and we use our fiducial estimate of $L_x$ (eqn.~\ref{eq:Lx}),
\begin{equation}
T_{\rm inner} \simeq 2.5 \times 10^5 L_{X,44}^{1/4} [(k/f)/0.08]^{-1/3} (M_*/M_\odot)^{-4/3 + 2\xi} M_{\rm BH,6.5}^{-2/3}\hbox{~K}.
\end{equation}
For fixed $L_x$, this is likely to be a lower bound on the temperature. To the degree that photon diffusion contributes to disk heat transport deep inside the flow, the lower surface density found toward smaller radii will lead to higher surface temperatures at small radii, concentrating the luminosity in a smaller area and raising its effective temperature.

If the ratio between the temperature near the ISCO to the temperature at $R_p$ changes slowly over time, whether it is controlled by dissipation of MHD turbulence or adiabatic compression, one might expect a slow decline in $L_X$ over time because the rate at which mass enters this region is roughly constant from 2--$3t_0$ until 9--$10t_0$, and the declining entry rate thereafter is partially compensated by the increasing ratio of inflow time to photon diffusion time.    In any event, there is little relation between its time-dependence and the $t^{-5/3}$ scaling of the debris stream mass-return rate.

The thermal X-ray radiation should be moderately beamed because the accretion flow, while geometrically fairly thick, is still noticeably flattened.   Diffusive photon flux will therefore preferentially travel normal to the orbital plane.   Similarly, magnetically-buoyant bubbles will preferentially rise in that direction because it should be roughly parallel to the net gravity (i.e., including the rotational contribution to the effective potential).   Thus, in the frame of the flow's orbital motion, the emergent intensity should be limb-darkened.    In addition, if the surface of the flow rises outward, as indicated by the simulation of \citet{Shiokawa+2015} in which $H\simeq 0.4R$ almost independent of $R$, its outer portion will very effectively block a large solid angle around the orbital plane.   Its Compton optical depth in the vertical direction remains large out to well beyond $a_{\rm min}$ \citep{Piran+2015}, so its integrated radial optical depth out to $R \sim a_{\rm min}$ is even larger.   If its scale height were exactly $0.4R$, this material would block at least 40\% of solid angle around the black hole, more if the photosphere lies higher than a single scale height above the plane.    For this reason, it is possible that it may be difficult for distant observers to see the thermal X-ray emission in a sizable fraction of events.    By contrast, the optical emission, which we argue is made at $R \sim a_{\rm min}$, should be seen from nearly all directions.

The data on TDEs other than ASASSN-14li, such as they are, are consistent with this picture, but the statistics are very poor: there are only three other apparently thermal TDEs with X-ray observations taken soon enough after the flare to be meaningful.  In one case (D23-H1: \citet{Gezari+2009}), there is an upper bound of $< 7 \times 10^{40}$~erg~s$^{-1}$ assuming no interstellar absorption.  Another (ASASSN-15oi: \citet{Holoien+16a}) was detected and, like ASASSN-14li, had a very soft spectrum, but its luminosity was approximately constant at $\simeq 5 \times 10^{42}$~erg~s$^{-1}$, making it uncertain whether this X-ray luminosity was associated with the TDE.  In the third case (PTF10iya: \citet{Cenko+12}), the peak $L_x \sim 10^{44}$~erg~s$^{-1}$, but the spectrum was hard enough that the authors reporting it suggest the X-rays may be from a jet we see from just outside the beam.

\section{Comparison to ASASSN-14li Observations}

\subsection{Optical/UV}
\label{sec:opticaluv}
As the estimates of the previous section have already made apparent, the conditions predicted by our model roughly match those seen in ASASSN-14li even for our fiducial parameters.    Its peak optical/UV luminosity, $6 \times 10^{43}$~erg~s$^{-1}$ when integrated over the best-fit blackbody spectrum, is only slightly smaller than our predicted fiducial maximum luminosity in this band, $\simeq 8 \times 10^{43}$~erg~s$^{-1}$.    The observed color temperature, $3.5 \times 10^4$~K, is likewise only a little bit below our fiducial maximum temperature, $\simeq 5 \times 10^4$~K.    In addition, our model predicts, at least in rough terms, that the optical/UV luminosity should follow the mass-return rate once the optical luminosity begins to fall, i.e., it should scale $\propto t^{-5/3}$ after peak brightness, in keeping with the observed lightcurve.   Note, however, that, unlike \citet{Miller+15}, we do {\it not} identify the time of peak optical output with the date of disruption; in our model, the peak is reached at $\simeq 7t_0$ after the disruption.    

Although we have not made specific predictions about which line features should be visible, the range of emission line widths (1700 -- 7700~km~s$^{-1}$ FWHM in the UV: \citet{Cenko+16}; initially $\simeq 10000$~km~s$^{-1}$ FW0I in the optical, but narrowing by a factor $\sim 2$ later in the event: \citet{Holoien+16}) is very consonant with an origin in a surface layer covering a flattened quasi-thermal surface at radii $\gtrsim a_{\rm min}$, where the circular orbital speed is $\simeq 10,000$~km~s$^{-1}$.    The fact that the emission lines' mean velocity shift with respect to the host galaxy is a small fraction of their width is consistent with the conditions near the outer shocks, in which the effective temperature at the flow surface varies slowly around a fluid element's orbit.  

The region responsible for the UV and X-ray absorption lines must be well separated from the optical/UV emission line region because the characteristic velocity of the absorbing material is an order of magnitude smaller than that of the emission line gas.  However, the fact that the absorption line profiles in the UV and X-ray are so similar \citep{Miller+15,Cenko+16} suggests that a single region accounts for both bands' absorption features.

It is also possible to use existing optical/UV data, both emission line profiles and light curves, to test other models for ASASSN-14li. In one model, the emission lines are radiated by the unbound debris \citep{Strubbe.Quataert.2009}.  As we have already argued in the context of the radio emission, the unbound debris are expelled over a particular range in directions that spans only a small fraction of the solid angle around the black hole.  Unless the mean direction of ejecta expulsion is very close to the sky plane, emission lines generated in the ejecta would have a mean velocity shift comparable to the line width, contrary to what is observed in ASASSN-14li.  Models containing an optically-thick expanding reprocessing region with a different origin \citep{Metzger.Stone.2015} have similar difficulty avoiding a mean shift comparable to the line width.  Emission lines primarily radiated from the illuminated side of an optically thick reprocessor could be seen only from receding material, and possibly not seen at all if near-side optically thick material lies on the line of sight.   Conversely, lines primarily radiated from the shadowed side of the reprocessor could be seen only from matter on the near, approaching side of the system.   In either case, there would be a sizable net shift.

The light curve itself also poses a significant constraint.  \citet{Alexander+16}, following the methods of \cite{Guillochon.Manukian.Ramirez-Ruiz2014}, find that the disruption began some time in spring 2014, the mass-return rate became super-Eddington in June or early July, reached a peak $\simeq 2.5$ in Eddington units in mid-September, and has been declining since then.  They further assume that the bolometric luminosity follows the mass-return rate, but is capped at Eddington.
However, \citet{Holoien+16} note that the optical flux on 13 July 2014 was at least a factor of 3 below the peak flux in mid-November ($m_V > 17$~mag as opposed to $m_V = 15.8$~mag).
If the bolometric luminosity reached a maximum in June that persisted until later than mid-September, for the optical flux in July to be more than a factor of 3 below that measured in mid-November requires a very sharp increase in the ratio of optical to bolometric luminosity beginning well after the peak in mass-return rate.  Because \citet{Miller+15} find that the ratio of V-band flux to X-ray flux steadily declines for at least 60~d post-discovery, the little evidence in hand does not support such an increase.  Although \citet{Alexander+16} assert that their model is consistent with the July upper bound, they reveal neither their most likely date of disruption, nor their favored decline rate after mid-September, nor any information about the relation between the optical and bolometric luminosities.

We further note that in the Alexander et~al. model, the peak in the mass-return rate was reached at least 120~d after disruption.  Simulations \citep{Guillochonweb,Shiokawa+2015} generally find that this peak occurs $\simeq 1.5t_0$ after disruption, implying $t_0 > 80$~d, a surprisingly large timescale.   Thus, the principal features of the lightcurve prior to discovery advocated in \citet{Alexander+16}---a disruption date in spring 2014 and a peak luminosity from mid-June or early July until mid-September 2014---are difficult to reconcile with the July 2014 upper bound.

\subsection{Soft X-rays}

In the soft X-ray band, our model predicts a lightcurve with a flattish peak stretching from 2--$3t_0$ to 8--$10 t_0$ followed by a gentle decline; in other words, we expect the X-ray flux to begin falling $\simeq 2t_0$ after the optical flux begins to fall.   At this somewhat vague level, our predicted lightcurve is consistent with the comparably uncertain shape of the X-ray lightcurve as determined by either \citet{Miller+15} or \citet{Charisi+16}.  In both analyses, the X-ray emission begins its decline $\simeq 20$--30~d after discovery, while the optical light declined monotonically after discovery. Our fiducial estimate of $L_X \sim 1.5 \times 10^{44}$~erg~s$^{-1}$
is about a factor of 2 below the observed value $3 \times 10^{44}$~erg~s$^{-1}$ \citep{Miller+15,Charisi+16}.   Although our nominal X-ray temperature $\gtrsim 2.5 \times 10^5$~K is a factor of $\simeq 2.5$ below the one observed, it is also estimated in a fashion that automatically makes it a lower bound.    Thus, our predictions for the scale of the X-ray luminosity and its characteristic temperature are at least approximately vindicated.    Our prediction for the timing relation between the optical emission and the X-ray is as consistent with the observations as it can be, given the uncertainties in the data.   

\subsection{Radio}\label{sec:radiodata}

Using our model to describe the radio data requires a longer discussion because, as shown in equations~\ref{nua} and \ref{Fnua}, it depends strongly on the density of the external gas and how it varies with distance from the black hole.   This density cannot be predicted within our model because its origin is wholly independent of the tidal disruption event, and is likely to vary substantially from galaxy to galaxy.   Here we will show that a modest amount of model-fitting to the observed data results in parameters easily consistent with our picture.

\citet{Alexander+16} report multi-frequency, multi-epoch radio observations of ASASSN-14li (see their Fig.~1). 
The observations began 2014 December 24 and continued until 2015 August-September. In their analysis, \citet{Alexander+16} subtract a possible quiescent AGN contribution from the observed signal.    For our purposes, we require a reduced form of their data, $\nu_a$ and $F_{\nu_a}$ as functions of time.   Because their data, though multi-frequency, is nonetheless taken at discrete frequencies, we show approximate values for these quantities in Table~\ref{table1}. We present both the ``corrected" data, from which a steady state signal has been subtracted and the uncorrected observations. 

\begin{table}[h]
\caption{Peak flux and peak frequency of the radio observations \citep[from][]{Alexander+16}.  Corrected values correspond to subtraction of a steady source, while uncorrected ones correspond to the observed values. }
\begin{center}
\begin{tabular}{|c|c|c|c|c|}
    \toprule
Date        &        $ \nu_{\rm peak}$ (10 GHz)  & $ F_{\rm peak}$ (mJy) &$\nu_{\rm peak}$ & $F_{\rm peak}$ (mJy)  \\
                                & corrected & corrected & & \\
\midrule
24 Dec. 2014           &   1--2                &        1.8--1.9 & 1.5 & 2 \\
6-13 Jan. 2015        &   0.8--1.3           &        1.8--1.9 & 0.9 & 2 \\ 
13  Mar. 2015          &    0.5                  &       1.2        & 0.4 & 2 \\ 
21-22 Apr.  2015     &    0.3--0.5           &       1.0        & 0.3 & 2\\ 
16-21 June 2015    &     0.2--0.3           &        0.9      & 0.2 & 2\\ 
28 Aug --1 Sept  2015    &     0.14--0.2  &       0.6--0.7&  0.14 & 2 \\
    \bottomrule
\end{tabular}
\end{center}
\label{table1}
\end{table}

Our analysis of this data proceeds in two steps.   First, using $\nu_a$ and $F_{\nu_a}$ at each of these epochs, we employ the equipartition formalism of \citet{Barniol+13} in order to infer the radius $R$ of the emitting region, the total number $N_e$ of relativistic electrons, and the magnetic field intensity $B$ in the emitting region.    The latter two quantities are determined as functions of $R$ by matching to $\nu_a$ and $F_{\nu_a}$; $R$ is then determined very tightly by the condition of minimizing the total energy $E_{eq}$ in relativistic electrons and magnetic field with respect to $R$, i.e., applying the equipartition condition.    In so doing, we assume (as did \citet{Alexander+16}) that the outflow is sub-relativistic and that the electron distribution function extends to low enough energies that the lowest characteristic synchrotron frequency radiated is less than $\nu_a$. Formally, of course, this analysis gives only a lower limit on the total energy.
 
\begin{figure*}[!ht]
\centering
\includegraphics[width=1\textwidth]{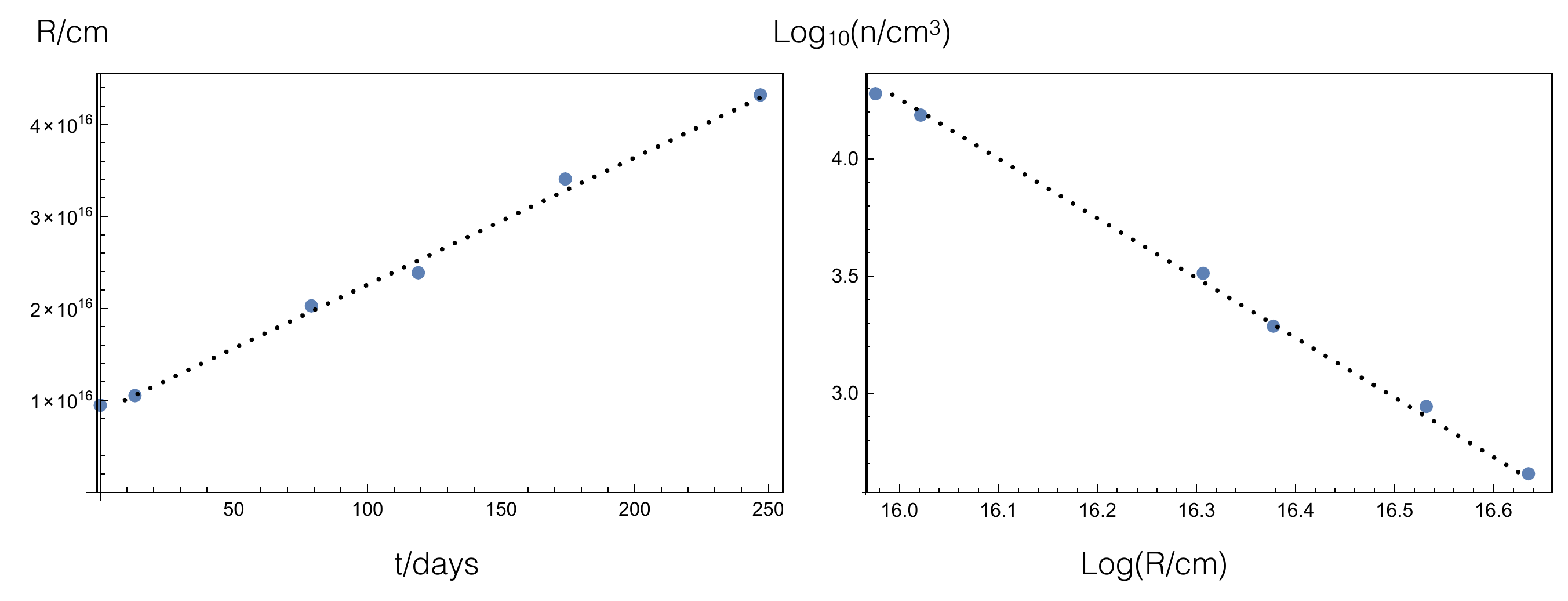}
\caption{The results of the equipartition analysis. Left: The emitting radius as a function of time since the first radio measurement. Right: ambient gas number density as a function of radius. The dashed lines depict a fit to the results that arise from independent analysis of the different observations. The regularity of the results (a constant velocity and a clear power-law decay of the density) support the validity of this model. }
\label{fig:Equipartition}
\end{figure*}

The first result of this procedure is to find that the emission radius increases from $8.5\times 10^{15} f_A^{-8/19} f_V^{-1/19} {\rm cm} $ at the first observation to $2.9 \times 10^{16}  f_A^{-8/19} f_V^{-1/19}   {\rm cm} $ at the last observation (see Fig.~\ref{fig:Equipartition}a).  Moreover, this increase is very nearly linear in time, corresponding to a nearly constant velocity $v \simeq 14,500 f_A^{-8/19} f_V^{-1/19}$~km~s$^{-1}$. This velocity is comparable to the expected velocity of the ejecta.  Thus, an analysis of the radio emission wholly without regard to the dynamics creating the relativistic electrons results in an outflow speed very similar to what would be expected for the unbound tidal debris.  If $f_A \sim 0.2$ in agreement with our bow shock estimate, the expansion speed might be a factor $\sim 2$ larger. However, as we show later, only a small fraction of the unbound mass is required to explain the observed radio luminosity. It is possible and even likely that such a small fraction of the ejecta would have somewhat greater energy than the nominal scale and therefore run ahead of the rest. We note here that the constancy of the velocity (see Fig.~\ref{fig:Equipartition}a) is remarkable given that the equipartition analysis is carried out for each point independently. {\it A priori}, there is no reason that this independent analysis should give such a simple, consistent relation between the different results. The resulting constant-speed expansion at a velocity comparable to $v_{\rm out}$ is a strong indication of the validity both of the equipartition analysis and of our model in which the outflow mass is much larger than the ambient mass it sweeps up.

 Our numerical results for $R(t)$ are very similar to those of \citet{Alexander+16}.    However, our interpretation is quite different: they suggested the outflow was driven by radiation pressure associated with the tidal flare, rather than identifying it with the unbound tidal debris. The equipartition analysis by itself cannot distinguish between the two models because it is wholly independent of the energy source for the emission: it only explores the conditions within the emitting region, obtaining a very robust estimate of the size of the emitting region and hence on the expansion velocity, and and somewhat weaker bounds on other quantities.
 

There are indications that our model is preferable. First, the robust velocity estimate agrees well with the predicted velocity of the unbound material. There is no {\it a priori} reason why a radiation pressure-driven outflow should have a speed so close to the ejecta speed. Second, if the bolometric luminosity was constant from June until after mid-September, why would the initiation of a radiation pressure-driven outflow be delayed three months after that maximum luminosity was reached?  In our model, however, the close agreement between the expansion velocity inferred from observations and the predicted velocity of unbound ejecta suggests strongly that the outflow began at the disruption; in that case, the disruption itself occurred in mid-September 2014, or about 70 d before optical discovery.  

We also note that this inferred timescale is extremely insensitive to the equipartition analysis because it is $R(t_{\rm max})/\langle dR/dt\rangle $, where $R(t) $ is the equipartition-derived radius as a function of time, $t_{\rm max}$ is the time of the final radio observation, and the angle brackets denote averaging. A systematic error in the scale of the inferred R cancels; all that remains is the timescale on which the inferred scale changes. As a result, there is very little difference between the lifetime we find for the expanding synchrotron source and the one found by \citet{Alexander+16}.

The estimated final equipartition energy, $ E_{eq}= 2.5  \times 10^{47} f_A^{-12/19} f_V^{8/19}$~erg, is only a lower limit on the total energy of the ejecta, but it is still more than three orders of magnitude smaller than the total ejecta energy.   Put another way, it also puts a lower bound on the ejecta mass that is quite small,
$M_{\rm min} = 2 E_{eq}/v^2 = 10^{-4} f_A^{4/19}  f_V^{10/19} M_\odot$.
 Such a small minimum mass is the justification for our earlier claim that the leading edge of the ejecta may travel a factor of several faster than the nominal expected speed.  It is also the justification for our earlier claim that the ejecta driving the shock came from a very thin layer at the surface of the star.

The external density, which follows from the number of electrons, is automatically comparable to $M_{\rm min}/m_p$ divided by the emitting volume\footnote{Like other estimates based on equipartition arguments, this density estimate is a lower limit. The density could be higher if the process is less efficient or if there is a significant number of non-relativistic electrons whose energy is insignificant in the total energy budget.}.  It decreases from $\approx 1.9 \times 10^4 f_A^{12/19} f_V^{-8/19}$~cm$^{-3}$ at a distance of $8.5\times10^{15} f_A^{-8/19} f_V^{1/19}$~cm to $\approx 450 f_A^{12/19} f_V^{-8/19}$~cm$^{-3}$  at $2.9 \times10^{16}f_A^{-8/19}  f_V^{1/19}$~cm.  The  decline is fit quite well by $n \propto r^{-2.5}$ (see Fig.~\ref{fig:Equipartition}b).  The regular power-law decline in the density is another indication supporting this analysis.  Just as for the successive radii, there is no reason that an independent analysis at different moments in time should result in such a smooth density profile.
This density is larger by about one order of magnitude than the corresponding density around the SMBH at our galactic center.  It is also slightly higher than that inferred for the TDE Swift~J1644 \citep{BarniolPiran2013}.   There is, of course, no reason to expect that conditions will be the same around different galactic center black holes, but the fact that the results  are comparable is reassuring.   
Note that, despite the unsupported assertion made by \citet{Alexander+16}, even at this density the surrounding material will not present any significant free-free opacity.  Integrating outward from the smallest inferred radius ($8.5 \times 10^{15}$~cm), the free-free optical depth is only $\simeq 0.03 T_4^{-3/2} \nu_{\rm GHz}^{-2}$.  We have deliberately scaled to a temperature of $10^4$~K in order to be conservative.  The first 30 days of the observed soft X-ray flux contain $\sim 10^7\times$ as many ionizing photons as there are electrons in the external medium out to the maximum radius inferred for the radio source, so essentially every atom should be stripped, and the electrons' characteristic energy will be $\sim 50$~eV, the temperature of the X-ray spectrum.   In such a state of high ionization, bremsstrahlung dominates the cooling rate; the associated cooling time is $\sim 200 (r/8.5 \times 10^{15}$~cm$)^{2.5}$~yr.   Thus, the temperature in the external gas is likely $\simeq 6 \times 10^5$~K, making the free-free optical depth even smaller.
 
 A clear prediction of our interpretation is that the radio source will continue to expand without slowing down.   Unfortunately, the radio emission is decreasing rapidly because of the drop in the external density, and it is not clear how long it will be detectable above the possible steady-state source in this galaxy. 

As we have emphasized, equipartition analysis is done without reference to the source of the energy.  If placed in the context of our earlier estimate based on the energy deposited in the ejecta's bow shock (eqns.~\ref{nua} and \ref{Fnua}), our results imply $\epsilon_e \sim \epsilon_B \sim 1$.   Alternatively, if $\epsilon_e \sim \epsilon_B \sim 0.1$ as in that earlier estimate, the ambient density would be roughly an order of magnitude greater.
 
\citet{vanVelzen+2015} suggested another interpretation, that the radio emission arose from a mildly relativistic jet.   To explore this possibility, we have also attempted to find a relativistic equipartition model that fits the data.  Following 
van~Velzen et~al., we assume that this relativistically-moving plasma accounts for the entire observed radio flux.  Because the van~Velzen et~al. data has limited spectral coverage, we use the ``uncorrected" 
\citet{Alexander+16} data for this analysis (the results change only slightly if we use the ``corrected" data in which a constant background source has been subtracted).    A relativistic solution can be found, but it requires a bulk Lorentz factor $\Gamma \approx 75  f_A^{-7/10} f_V^{1/17} $ with no deceleration.  The required Lorentz factor would be $\approx 220$ if we used the ``corrected" data.  However, the external shocked mass needed to produce the observed radiation implies that such a relativistic outflow would have decelerated substantially during the time of the observations.  Alternatively, one can reduce $\Gamma$ significantly, but only at the price of positing an emission region that is far from equipartition, and therefore requires considerably more than the minimum energy.   For example, to obtain $\Gamma \sim 2 $ instead of 75, one needs an increase by a factor of  $\sim 10^8$ in the total energy, implying a required energy of $> 10^{53}$~erg in the jet.

\vskip 0.5cm

\vskip 0.5cm

To summarize this section, we are able to reproduce quite well the observed radio emission if there is a time-steady component as posited by \citet{Alexander+16}.    The expansion speed of the radio source matches the predicted outflow speed of the unbound tidal debris, and both the shape and amplitude of the radio spectrum are reproduced with a very plausible external density profile.  Thus, we suggest that the observed radio signal was produced by the unbound ejecta, and there is no need to invoke an additional component to produce it.

\subsection{Characteristic timescale}
\label{sec:timescale}

The final step in comparison of our model to ASASSN-14li is to use the relative timing of the different components to estimate the characteristic timescale $t_0$.    We have just argued that the radio data suggest the tidal disruption took place $\simeq 70$~d before discovery.   If the observed optical light curve were extrapolated to earlier times as $(t-t_d)^{-5/3}$, one would infer a peak $\simeq 35$~d before discovery \citep{Miller+15}.   However, there is no optical data during that time, so we do not know whether such an extrapolation is appropriate.   If we instead suppose that the optical peak either coincided with discovery or happened earlier, and follow our model in which the optical emission begins to decline at $\simeq 7t_0$, the implied $t_0$ is $\gtrsim 10$~d.

The X-ray light curve can be used in a similar way because our model indicates that the X-ray flux begins to diminish around 8--$10t_0$.  The X-ray data analysis of \citet{Miller+15} and \citet{Charisi+16} shows the beginning of the decline to occur $\simeq 20$--30~d after discovery, pointing to $t_0 \approx 9$--12~d, in excellent agreement with thelower end of the range implied by our analysis of the optical lightcurve.    We emphasize that within our model the optical and X-ray estimates of $t_0$ are physically quite independent.  One depends on events at $R \sim a_{\rm min}$, the other on events at $R \sim R_p$; their close agreement is therefore by no means built-in.

Thus, our model applied to both the X-ray and optical light curves suggests a $t_0$ roughly half our fiducial value.   Taken at face-value, this result implies a geometric mean of the black hole and stellar mass about $3/4 \times 10^6 M_{\odot}$.  However, we caution that because $t_0 \propto (k/f)^{1/2}$, it is also sensitive to the star's internal structure; a genuine calculation of the internal structure of main sequence stars of near-solar mass might yield a value of $k/f$ that changes our fiducial estimate of 20~d by a factor of order unity.


\section{Summary}

All but one of the many quantitative predictions made by our model (optical luminosity, temperature, timescale, line-widths; X-ray luminosity, temperature, and timescale; radio spectrum and inferred expansion speed) matches the measured properties of ASASSN-14li to within a factor of 2 or even closer. The largest nominal discrepancy (a factor that is still only  $\sim 3$, even taking the conservative assumption $f_A \sim 0.2$) is with the predicted radio expansion speed, but this higher velocity can be attributed to a small tail in the unbound mass's energy distribution.
This level of agreement is consistent with the combined uncertainties of the predictions and the data.   Our predictions also match the shape of the observed time-dependence in all three bands at the level of the uncertainty in the data.   All this is done with only two free parameters, the external gas density and its logarithmic radial derivative, and these parameters are well within the range one might expect.    We argue that this extraordinary degree of quantitative matching provides strong support for this picture. At the same time, however, we recognize that we have not explained every property of this source; in particular, interpreting the narrowing of the optical/UV emission lines over time is likely to require attention to the subtleties of how line formation in the flow's atmosphere depends on the local optical depth and heating rate.

We have also shown that certain features of other proposed models can, thanks to the specificity of the ASASSN-14li data, be ruled out.   For example, the lack of a significant mean velocity shift in the emission lines rules out an origin for them in either the unbound debris or a rapidly-expanding reprocessing region.    Likewise, the upper limit on the July flux from ASASSN-14li strongly undermines the case for a model in which the disruption took place early in 2014 and the radio-emitting outflow emerged only when the luminosity grew large enough to expel a sizable amount of matter.

Unfortunately, at the moment ASASSN-14li is the only TDE for which light curves are available in all three bands, optical, X-ray, and radio.  Thus, it is at present the only example in which all parts of this model can be tested.   However, at the current pace of discovery of new TDEs we are optimistic that additional examples will soon become available.   In addition, the success of the optical/UV portion of the model in reproducing the characteristics of seven TDEs with good observations in that band \citep{Piran+2015} is a promising sign of its general applicability.

Lastly, we would like to emphasize a purely empirical argument touching on a key point in the conventional view of TDE dynamics. As we have already pointed out, the time-dependence in ASASSN-14li of the largest contributor to the bolometric luminosity, the soft X-rays, observed for seven months after discovery, bears no resemblance to $t^{-5/3}$.   This observation is of central importance because nearly all earlier analyses of TDEs have assumed that the bolometric output should decline in proportion to this power of the time since disruption once the mass-return rate peaks, an event estimated to occur $\simeq 1.5t_0$ after disruption.    For such a time-dependence to apply, most of the matter destined to be accreted by the black hole must have its orbit ``circularized" at radii not too far outside $R_p$, and do so within a time $\lesssim t_0$ of its first return to the pericenter region, so that the energy release can track the mass-return rate.    If the bolometric luminosity does {\it not} follow the expected proportionality to $t^{-5/3}$, the effort to discover rapid circularization mechanisms is unnecessary.

\acknowledgments
We thank Rodolfo Barniol Duran, Ehud Nakar, Re'em Sari,   Elad Steinberg and Almog Yalinewich  for helpful discussions.
Two of us (JK and RC) would like to thank Hebrew University and its Institute for Advanced Studies for hospitality while the key ideas of this paper were developed.   We are also grateful to Ehud Nakar for a loan of office space.   This work was supported by the grants:  I-CORE  1829/12,  ISA  3-10417  (TP);  NSF  AST-1028111 and AST-1516299, and NASA/ATP  NNX14AB43G (JK).

\bibliographystyle{apj}
\bibliography{tde}

\end{document}